\documentclass[a4paper,12pt]{article}
\pdfoutput=1
\usepackage{authblk}
\usepackage{amsmath}
\usepackage{multirow}
\usepackage[margin=1.25in]{geometry}
\usepackage{natbib}

\def \bbeta {\boldsymbol{\beta}}
\def \betaa {\boldsymbol{\eta}}
\def \bgamma {\boldsymbol{\gamma}}
\def \btheta {\boldsymbol{\theta}}
\def \bpsi {\boldsymbol{\psi}}

\def \bQ {\boldsymbol{Q}}
\def \bX {\boldsymbol{X}}

\def \bh {\boldsymbol{h}}
\def \bq {\boldsymbol{q}}

\def \bx {\boldsymbol{x}}

\def \MARt {MAR$_T$}
\def \MARtx {MAR$_{T,\bX}$}

\def \MARtxq {MAR$_{T,\bX,\bQ}$}
\def \MARq {MAR$_{\bQ}$}
\def \sMARt {MAR$_T$ }
\def \sMARtx {MAR$_{T,\bX}$ }

\def \sMARtxq {MAR$_{T,\bX,\bQ}$ }
\def \sMARq {MAR$_{\bQ}$ }

\DeclareMathOperator{\expit}{expit}

\def\bSig\mathbf{\Sigma}

\usepackage[figuresright]{rotating}
\title{The competing risks Cox model with and without auxiliary case covariates under weaker or no missing-at-random cause of failure}

\author[1,2]{Daniel Nevo\thanks{danielnevo@gmail.com}}
\author[3,4]{Reiko Nishihara}
\author[2,4,5]{Shuji Ogino}
\author[1,2]{Molin Wang}

\affil[1]{\small Department of Biostatistics, Harvard T.H. Chan School of Public Health}
\affil[2]{\small Department of Epidemiology, Harvard T.H. Chan School of Public Health}
\affil[3]{\small Department of Nutrition, Harvard School of Public Health}
\affil[4]{\small Department of Medical Oncology, Dana-Farber Cancer Institute}
\affil[5]{\small Division of MPE Molecular Pathological Epidemiology, Department of Pathology \\ Brigham and Women's Hospital and Harvard Medical School}

\date{}
\begin{document}




\label{firstpage}


%
%

\newgeometry{top=1in,bottom=1in,right=1.25in,left=1.25in}

\maketitle
\vspace{-1cm}
\begin{abstract}
In the analysis of time-to-event data with multiple causes using a competing risks Cox model, often the cause of failure is unknown for some of the cases. 
The probability of a missing cause is typically assumed to be independent of the cause given the time of the event and covariates measured before the event occurred.  In practice, however, the underlying missing-at-random assumption does not necessarily hold. Motivated by colorectal cancer subtype analysis, we develop semiparametric methods to conduct valid analysis, first when additional auxiliary variables are available for cases only. We consider a weaker missing-at-random assumption, with missing pattern depending on the observed quantities, which include the auxiliary covariates. Overlooking these covariates will potentially result in biased estimates. We use an informative likelihood approach that will yield consistent estimates even when the underlying model for missing cause of failure is misspecified. We then consider a method to conduct valid statistical analysis when there are no auxiliary covariates in the not missing-at-random scenario.  The superiority of our methods in finite samples is demonstrated by simulation study results. We illustrate the use of our method in an analysis of colorectal cancer data from the Nurses' Health Study cohort, where, apparently, the traditional missing-at-random assumption fails to hold for particular molecular subtypes.  
\end{abstract} 
\restoregeometry
\section{Introduction}
\label{Sec:intro}
Disease heterogeneity is  commonly represented by molecular subtyping of the disease. When considering time-to-disease diagnosis data, each of the disease subtypes can be considered as a cause for the event of interest, the disease. Those subtypes, typically defined by biomarkers, are used for better understanding of  biological mechanisms behind the disease. More recently, scientists have found that  influences of disease  risk factors can vary across disease subtypes in molecular pathological epidemiology (MPE) research \citep{ogino2016role}. For example, the effect of some life-style factors on the risk of colorectal cancer (CRC) is different across CRC subtypes \citep{ogino2010lifestyle,campbell2010case}. Statistical methodology has been developed to evaluate this type of etiologic heterogeneity across disease subtypes \citep{chatterjee2010analysis,begg2013conceptual,wang2016statistical,nevo2016}.

However, subtype data are often missing for some of the disease cases due to various reasons. For tumor tissue data, for example,   biomarkers defining the tumor subtypes may be missing because tumor tissue is unavailable, or because collected tumor tissue  is insufficient in quantity of quality. One commonly-used approach for covariates effect estimation in the proportional hazard  (PH) competing risks models with missing cause of failure, originally suggested by \cite{goetghebeur1995analysis}, is to construct estimating equations emerging from two separate partial likelihoods \citep{goetghebeur1995analysis,dewanji1992note}.  These two partial likelihood approaches were compared and carefully studied by \cite{lu2005comparison}. More recently, \cite{chatterjee2010analysis} suggested a two-stage model for the scenario  where a large number of subtypes are composed from multiple markers, and  generalized the estimating equation method of \cite{goetghebeur1995analysis} to estimate regression parameters under their two-stage model. 

For this missing cause of failure problem, other methods developed include  parametric methods   \citep{flehinger2002parametric}, Bayesian methods \citep{basu2003bayesian,sen2010bayesian}, as well as implementation of commonly used missing data methodology, such as multiple imputation \citep{lu2001multiple}, utilization of the EM algorithm \citep{craiu2004inference,craiu2006inference}, and inverse probability weighting (IPW) approach  \citep{gao2005semiparametric,hyun2012proportional}. Most of the aforementioned methods  assume a missing-at-random (MAR) assumption that, given the event time and the covariates measured  for all subjects, including subjects with and without the disease of interest, the probability of a missing cause is the same for all cases. 

Often, the MAR assumption is invalid in applications. For example, it is possible that data on   biomarkers, and therefore subtypes, are more likely to be missing when a tumor exhibits certain characteristics. In our motivating example, where of interest are the effects of risk factors on CRC subtypes microsatellite instability (MSI) and microsatellite stability (MSS) \citep{cancer2012comprehensive,kocarnik2015molecular},  the MSI/MSS status is missing for more than half of the cases. Our data analysis suggested that CRC tumor location (proximal colon, distal colon or rectum), which is measured for all cases we considered, may be indicative of the existence of adequate tumor tissue for obtaining the subtype data. Furthermore, it is well established in the subject-matter literature that CRC tumor location is associated with MSI/MSS status \citep{colussi2013molecular}.   In short, one goal of this paper is to provide methodology taking advantage of the information about missingness contained in some disease characteristics (for example, tumor location in our motivating example). When we do not have information about disease characteristics that are associated with the missingness status, we will also propose a method that does not require the MAR assumption; this is the second goal of the paper. While there exist methods that do not require MAR, they typically assume the missing status does not depend on event time or covariates \cite[e.g.,][]{flehinger2002parametric}.



The rest of the paper is organized as follows. In  Section \ref{Sec:model}, we present PH models for competing risks, allowing for missing subtypes, and discuss various assumptions concerning the missingness pattern. In Section \ref{Sec:Estim}, we describe our approach in the presence of auxiliary case covariates which is valid under a weaker MAR assumption. In Section \ref{Sec:NMAR}, we present a method  allowing for not missing-at-random (NMAR) subtype data.    Finite sample properties of the suggested  methods are investigated using simulations in Section \ref{Sec:Sims}. Analysis of our motivating example, the CRC subtype data, is presented in Section \ref{Sec:Data}.  Concluding remarks are given in Section \ref{Sec:Discuss}. 

\section{Models and notations}
\label{Sec:model}
We consider time-to-event data within the competing risks framework.  Let $\tilde{T}$ be the time until the occurrence of the disease and $Y$ be the subtype. $Y$ takes a single value from $\{1,...,K\}$. 
Denote also $\bX=\bX(t)$ for a vector of possibly time-dependent covariates. A Cox-type model for the subtype-specific hazard is then
\begin{equation}
\label{Eq:HazardDef}
\lambda_k(t):=\lim\limits_{\Delta t\rightarrow0}\frac{1}{\Delta t}P(t\le \widetilde{T}<t+\Delta t, Y = k|\widetilde{T}\ge t,\bX=\bx)=\lambda_{01}(t)\alpha_k(t;\betaa)\exp(\bbeta^T_k\bx)
\end{equation}
with $\lambda_{01}(t)$ being an unspecified baseline hazard function for subtype $1$ and where $\alpha_k(t;\betaa)=\lambda_{0k}(t)/\lambda_{01}(t), k=1,...,K,$  are baseline hazard ratio functions; $\alpha_1(t)=1$ by definition. This model allows an unspecified baseline hazards, as in the usual Cox model \citep{cox1972regression}, while allowing the baseline hazard ratio between two subtypes to depend on $t$, possibly via a parametric model.    If the number of different subtypes is not too large, a possible approach is to specify a piecewise constant function for $\alpha_k(t;\betaa)$, for each $k$. While we operate under the commonly used proportional hazard model, the  methodology we present  can be extended to a broader set of regression functions.  Let $C$ be the censoring time,  $T=\min(\widetilde{T},C)$, the observed time, and  $\delta=I\{T=\widetilde{T}\}$, the disease censoring indicator. We assume that given covariates, censoring time is independent of both $\tilde{T}$ and $Y$ \citep{kalbfleisch2011statistical}. 


The model described above is extended to allow missing subtype data by introducing an indicator $O$ that equals to one if the subtype $Y$ is observed, and zero otherwise.  If $\delta=0$, then $O=0$ as well. Denote 
\begin{equation}
\label{Eq:MissProbNoQ}
\pi(t, \bx, k):=P(O=1|\tilde{T}=t, \bX=\bx, Y=k)
\end{equation} 
for the probability of observing the subtype given a diagnosis at time $t$, for a subject with a covariates vector $\bx$ and subtype $k$.  \cite{goetghebeur1995analysis} considered a model where the missing subtype probability depends on the event time $(\tilde{T})$ only. That is, $\pi(t, \bx, k)=\pi(t)=P(O=1|\tilde{T}=t)$.  This is a form of a MAR assumption that we denote by \MARt. Similarly, we denote \sMARtx for the commonly used assumption $\pi(t, \bx, k)=\pi(t, \bx)=P(O=1|\tilde{T}=t, \bX=\bx)$. We also denote $\overline{\pi}=1-\pi$ as the probability of a missing subtype. 

Often, in practice, subtypes (i.e., causes) are more likely to be missing when the disease exhibits specific characteristics which are associated with the subtypes, even after controlling for $\bX$ and $\tilde{T}$. In these situations, the missingness probability depends on the subtype even after controlling for $\bX$ and $\tilde{T}$; that is, neither \sMARtx nor \sMARt are expected to hold. However, often, it is reasonable to assume that the missing probability is independent of the subtype status when conditioning on these disease characteristics mentioned above in addition to $\bX$ and $\tilde{T}$; this is a weaker assumption than \sMARtx and \MARt.
We denote by $\bQ$ the vector of measured disease characteristics,   which we assume is available for all cases,   and we denote its joint probability function by 
$$\nu_k(\bq, \bx, t;\bpsi)=P_{\bpsi_k}(\bQ=\bq|Y=k,\bX=\bx,\tilde{T}=t),
$$
with $\bpsi_k$ being an unknown vector of parameters.  In practice, $\bQ$ might be  conditionally independent of  either $T$, $\bX$ or both,  given $Y$. Note that $\bQ$ may include both discrete and continuous variables. It can also be a single random variable, as in the example presented in Section \ref{Sec:Data}.  In that example, $Q$ is the tumor location and it is possibly associated with the probability of missing CRC subtype; see Section \ref{Sec:Data}.  

To account for $\bQ$ effect on the possibility of missing subtype, the probability of observing the subtype, previously defined in \eqref{Eq:MissProbNoQ}, is redefined   as
\begin{equation}
\label{Eq:piDef}
\pi(t, \bx, \bq, k):=P(O=1|\tilde{T}=t, Y=k, \bX=\bx, \bQ=\bq),
\end{equation}
which is written as $\pi(t, \bx, \bq, k;\bgamma)$ when a model with a vector of parameters $\bgamma$ is assumed. We can now introduce  more relevant possible assumptions regarding the missing mechanism:
\begin{align}
\pi(t,  \bx, \bq, k)& = \pi(t, \bx, \bq)  = P(O=1|\tilde{T}=t, \bX=\bx, \bQ=\bq) \tag{\text{\MARtxq\hspace{-2pt}}}, \\
\pi(t, \bx, \bq, k)& = \pi(\bq)        =   P(O=1|\tilde{T}=t,  		    \bQ=\bq) \tag{\text{\MARq}}.
\end{align}
Clearly, \sMARtxq is a weaker assumption  than \MARtx.  While subject-matter knowledge can help to determine which missing assumption is reasonable in a specific application, these assumptions cannot be tested using the observed data, as often the case when dealing with assumptions regarding missing data. 

When using the existing methods, described in Section \ref{Sec:intro}, the validity of analysis conducted relies on  correctness of the \sMARtx assumption as well as on  accurate specification of the parametric model $\pi(t,\bx;\bgamma)$.  Our methodology, presented in the next section, alleviate these concerns while replacing them with a assumptions on the conditional distribution of $\bQ$ given $Y,\tilde{T}$ and $\bX$.

\section{Estimation and inference under weaker MAR}
\label{Sec:Estim}
We now move to construction of a partial likelihood for estimation of the model parameters  $\betaa=\{\betaa_1,...,\betaa_k\}, \bpsi=\{\bpsi_1,...,\bpsi_K\}$ and $\bbeta=\{\bbeta_1,...,\bbeta_k\}$, although $\bbeta$ is of our main concern.  Under \MARtxq, we may now write (for all $t, \bx$ and $\bq$) 
{
	\begin{align}
	\nonumber
	& \lim\limits_{\Delta t\downarrow 0}P(t\le \tilde{T} < t + \Delta t, Y = k, \bQ = \bq, O = 1|\tilde{T}\ge t, \bX=\bx) \\ \nonumber
	&  = P(O=1| \tilde{T}=t, \bX=\bx, \bQ=\bq)\times P(\bQ=\bq|Y=k, \tilde{T}=t, \bX=\bx, )\\ \nonumber
	&\times\lim\limits_{\Delta t\downarrow 0}P(t\le \tilde{T} < t + \Delta t, Y = k|\tilde{T}\ge t,\bX=\bx) \\ 
	&  = \pi(t,\bx, \bq)\nu_k(\bq,\bx,t;\bpsi)\lambda_{01}(t)\alpha_k(t;\betaa)\exp(\bbeta^T_k\bx) \label{Eq:HazObs}
	\end{align}}
if the subtype $Y$ is observed, and 
\begin{align}
\nonumber
& \lim\limits_{\Delta t\downarrow 0}P(t\le \tilde{T} < t + \Delta t,  \bQ = \bq, O = 0|\tilde{T}\ge t,\bX=\bx) \\ \nonumber
&  = \sum_{k=1}^{K}\lim\limits_{\Delta t\downarrow 0}P(t\le \tilde{T} < t + \Delta t, Y = k, \bQ = \bq, O = 0|\tilde{T}\ge t, \bX=\bx) \\
\label{Eq:HazMiss}
&  = \sum_{k=1}^{K}\overline{\pi}(t, \bx, \bq)\nu_k(\bq,\bx,t;\bpsi)\lambda_{01}(t)\alpha_k(t;\betaa)\exp(\bbeta^T_k\bx)
\end{align}
if the subtype is not observed. 

Construction of partial likelihood for the PH competing risks model with missing subtypes for part of the cases entails some complexity. To present existing methods, we first consider the situation where $\bQ$ is not observed and is not part of the model. 
For the PH competing risks model, the standard approach when subtypes are observed for all cases  is to condition not only on the case occurrence, as in the standard Cox model with a single type of event, but also on its subtype.  Then, in a model without $\bQ$, the baseline hazard in the denominator and the numerator cancel out, $\pi_i=1$ for all $i$, and the resulting partial likelihood for $\bbeta$ is 
\begin{equation*}
\label{Eq:FullLik}
L(\bbeta)=\prod_{i=1}^{n}\left[\frac{\exp(\bbeta^T_{y_i}\bx_i)}{\sum_{j=1}^{n}\xi_j(t_i)\exp(\bbeta^T_{y_i}\bx_j)}\right]^{I\{\delta_i=1\}}
\end{equation*}
\citep{prentice1978analysis}, where $\xi_j(t)$ equals to one if $j$ is in the risk set at time $t$ and zero otherwise and $y_i$ is the observed subtype.

When subtypes are unobserved for some cases, two partial likelihood approaches, which differ in the conditioned event, have been discussed, typically under \MARtx, or stronger assumptions.  The first \citep{goetghebeur1995analysis} is to condition on the occurrence of an event, the observed subtype and on the fact that subtype was observed, and for events with unobserved subtypes, to condition on the occurrence of an event and on not observing the corresponding subtype. This is done by considering the hazard at time $t$ of event occurrence of any subtype,  which is the sum of the hazard functions for all subtypes, also at time $t$. That is,
$$
P(t\le \widetilde{T}<t+\Delta t|\widetilde{T}\ge t,\bX=\bx)= \sum_{k=1}^{K}P(t\le \widetilde{T}<t+\Delta t, Y = k|\widetilde{T}\ge t,\bX=\bx).
$$
Then, incorporating the missingness model under  \sMARtx and by using calculations similar to  \eqref{Eq:HazObs} and \eqref{Eq:HazMiss}, this likelihood \citep{goetghebeur1995analysis} becomes
\begin{equation}
\begin{split}
\label{Eq:L}
L(\bbeta,\betaa,\bgamma)&=\left\{\prod_{i=1}^{n}\left[\frac{\pi(t_i,\bx_i;\bgamma)\exp(\bbeta^T_{y_i}\bx_i)}{\sum_{j=1}^{n}\xi_j(t_i)\pi(t_i,\bx_j;\bgamma)\exp(\bbeta^T_{y_i}\bx_j)}\right]^{I\{\delta_i=1,O_i=1\}} \right.
\\
&\times\left.\left[\frac{\overline{\pi}(t_i,\bx_j;\bgamma)\sum_{k=1}^{K}\exp(\bbeta^T_{k}\bx_i)}{\sum_{j=1}^{n}\xi_j(t_i)\overline{\pi}(t_i,\bx_j;\bgamma)\sum_{m=1}^{K}\alpha_m(t_i;\betaa)\exp(\bbeta^T_{m}\bx_j)}\right]^{I\{\delta_i=1,O_i=0\}}\right\}.
\end{split}
\end{equation}
However, when \sMARtx does not hold, this likelihood cannot be used under the weaker \sMARtxq assumption since $\pi$ could not be calculated even when $\bQ$ is observed. This is because $\pi$ could not be calculated without knowing the value of $\bQ$ for subjects in the risk set with $\delta=0$.

We now construct a more informative partial likelihood, similar to the one suggested by \cite{dewanji1992note}. We condition on the fact that event was observed, but not on the event subtype. Taking this approach, we can write down the partial likelihood for the observed data $(T_i,\delta_i,\bX_i,Y_iO_i,O_i,\bQ_i\delta_i)$. For censored observations, $O_i$ is defined as zero and $Y_i$ and $\bQ_i$ are taken to be missing.   Following  \eqref{Eq:HazardDef},\eqref{Eq:HazObs} and \eqref{Eq:HazMiss} and assuming \MARtxq, we have 
\begin{align}
\label{Eq:LstarQ}
\begin{split}
L^\star_Q(\bgamma,\bbeta,\betaa,\bpsi)&=\prod_{i=1}^{n}\left\{\left[\frac{ \pi(t_i,\bx_i, \bq_i;\bgamma)\nu_{y_i}(\bq,\bx,t;\bpsi)\alpha_{y_i}(t_i;\betaa)\exp(\bbeta^T_{y_i}\bx_i)}{\sum_{j=1}^{n}\xi_j(t_i)\sum_{m=1}^{K}\alpha_{m}(t_i;\betaa)\exp(\bbeta^T_{m}\bx_j)}\right]^{I\{\delta_i=O_i=1\}}\right.\\
&\times \left.\left[\frac{\sum_{k=1}^{K} \overline{\pi}(t_i,\bx_i, \bq_i;\bgamma)\nu_k(\bq,\bx,t;\bpsi)\alpha_{k}(t_i;\betaa)\exp(\bbeta^T_{k}\bx_i)}{\sum_{j=1}^{n}\xi_j(t_i)\sum_{m=1}^{K}\alpha_{m}(t_i;\betaa)\exp(\bbeta^T_{m}\bx_j)}\right]^{I\{\delta_i=1,O_i=0\}}\right\}.
\end{split}
\end{align}
Note that when using $L^\star_Q$, the missing model need not to be correctly specified, as long as at least \sMARtxq holds. This is because we can decompose $L^\star_Q$ into $L^\star_{Q1}\times L^\star_{Q2}$, where
\begin{align}
\label{Eq:LstarDecomp}
\begin{split}
L^\star_{Q1}(\bgamma)&=\prod_{i=1}^{n}\left[(\pi(t_i,\bx_i, \bq_i;\bgamma))^{I\{\delta_i=O_i=1\}}\overline{\pi}(t_i,\bx_i, \bq_i;\bgamma)^{I\{\delta_i=1,O_i=0\}}\right],\\
L^\star_{Q2}(\bbeta,\betaa) & = \prod_{i=1}^{n}\left\{\left[\frac{\nu_{y_i}(\bq,\bx,t;\bpsi)\alpha_{y_i}(t_i;\betaa)\exp(\bbeta^T_{y_i}\bx_i)}{\sum_{j=1}^{n}\xi_j(t_i)\sum_{m=1}^{K}\alpha_{m}(t_i;\betaa)\exp(\bbeta^T_{m}\bx_j)}\right]^{I\{\delta_i=O_i=1\}}\right.\\
&\times \left.\left[\frac{\sum_{k=1}^{K}\nu_k(\bq,\bx,t;\bpsi)\alpha_{k}(t_i;\betaa)\exp(\bbeta^T_{k}\bx_i)}{\sum_{j=1}^{n}\xi_j(t_i)\sum_{m=1}^{K}\alpha_{m}(t_i;\betaa)\exp(\bbeta^T_{m}\bx_j)}\right]^{I\{\delta_i=1,O_i=0\}}\right\}.
\end{split}
\end{align}
Estimation of $\bbeta$ can be done based solely on $L^\star_{Q2}$.   The approach of \cite{goetghebeur1995analysis}, which is also utilized in \cite{chatterjee2010analysis}, is to combine the score function for $\bbeta$ from $L$ in \eqref{Eq:L} and the score function for $\betaa$ from  a likelihood $L^\star$, a likelihood analogue of $L^\star_Q$ in \eqref{Eq:LstarQ} for a model without $\bQ$, to form a system of  estimating equations. However, as we noted before, $L$ cannot be used when the missing probability depends on $\bQ$. It is worth mentioning that $\nu_k(\bq,\bx,t;\bpsi)$ cannot be estimated from the observed subtype data only since, if $O$ and $\bQ$ are conditionally dependent,
$$
\nu_k(\bq,\bx,t;\bpsi)=P(\bQ|\bX,T;\bpsi)\ne P(\bQ|O=1,\bX,T;\bpsi,\bgamma).
$$
Let $\btheta=(\bbeta,\betaa, \bpsi)$ and let  $\btheta_0$  be its true value. We suggest to estimate $\btheta$ by maximizing $L^\star_{Q2}$ (or $\ell^\star_{Q2}=\log L^\star_{Q2}$). This could be done using a standard Newton-Raphson algorithm. Denote $\hat{\btheta}$ for the estimated vector of parameters. It can be shown that $\hat{\btheta}$ is a consistent estimator.  Furthermore, $\sqrt{n}(\hat{\btheta}-\btheta_0)$ converges in distribution to a multivariate normal distribution with a covariance matrix that can be consistently estimated by a robust sandwich estimator. Details and proofs are given in Web Appendix A. 

The \sMARtxq assumption used for our method above is weaker than the often assumed \MARtx.  While it should be feasible to collect additional data on cases only, finding $\bQ$ such that  \sMARtxq exactly holds might be impossible in many applications. Even when both \sMARtx and \sMARtxq fail, one may expect that as long as $\bQ$ and $Y$ are associated conditionally on $T$ and $X$,  using our method would produce results less biased than those obtained by analysis under \MARtx, as $\bQ$ contains information about the unobserved value of $Y$.  The results of our simulations in Section \ref{Sec:Sims} considering the scenario under NMAR   provide evidence to this claim. In the next section, we present a method which is valid under NMAR, that can be used if there is no relevant $\bQ$ available.  
\section{Estimation and inference under NMAR}  
\label{Sec:NMAR}
In this section we consider the scenario where a $\bQ$ cannot be found such that \sMARtxq holds. The subtypes are considered to be  NMAR, that is, none of the standard MAR assumptions holds, and the probability of missing subtype does depend on the true subtype status. Similarly to  \eqref{Eq:HazObs} and \eqref{Eq:HazMiss}, we may write
\begin{equation*}
\lim\limits_{\Delta t\downarrow 0}P(t\le \tilde{T} < t + \Delta t, Y = k,  O = 1|\tilde{T}\ge t, \bX=\bx)  = \pi(t, \bx, k)\lambda_{01}(t)\alpha_k(t;\betaa)\exp(\bbeta^T_k\bx)
\end{equation*}
if the subtype $Y$ is observed, and 
\begin{align*}
\nonumber
& \lim\limits_{\Delta t\downarrow 0}P(t\le \tilde{T} < t + \Delta t, O = 0|\tilde{T}\ge t,\bX=\bx) = \sum_{k=1}^{K}\overline{\pi}(t,   \bx, k)\lambda_{01}(t)\alpha_k(t;\betaa)\exp(\bbeta^T_k\bx)
\end{align*}
if the subtype is not observed.  Similarly to the way we  formed $L^\star_Q$, we may write the partial likelihood obtained by conditioning on  the event occurrence but not on its subtype. This leads to 
\begin{align}
\begin{split}
L^\star_Y(\bgamma,\bbeta,\betaa)&=\prod_{i=1}^{n}\left\{\left[\frac{ \pi(t_i,\bx_i,  y_i;\bgamma)\alpha_{y_i}(t_i;\betaa)\exp(\bbeta^T_{y_i}\bx_i)}{\sum_{j=1}^{n}\xi_j(t_i)\sum_{m=1}^{K}\alpha_{m}(t_i;\betaa)\exp(\bbeta^T_{m}\bx_j)}\right]^{I\{\delta_i=O_i=1\}}\right.\\
&\times \left.\left[\frac{\sum_{k=1}^{K} \overline{\pi}(t_i, \bx_i, k;\bgamma)\alpha_{k}(t_i;\betaa)\exp(\bbeta^T_{k}\bx_i)}{\sum_{j=1}^{n}\xi_j(t_i)\sum_{m=1}^{K}\alpha_{m}(t_i;\betaa)\exp(\bbeta^T_{m}\bx_j)}\right]^{I\{\delta_i=1,O_i=0\}}\right\}.
\end{split}
\end{align}
Estimates for $\{\bgamma,\bbeta,\betaa\}$ are obtained by maximizing $L^\star_Y$. The resulting estimators are consistent and normally distributed, with a variance that could be consistently estimated. Unlike $L^\star_Q$, $L^\star_Y$ cannot be partitioned and thus the parameters in $\pi(t, x, y;\bgamma)$ must be estimated. While this method is valid, and could be used under NMAR, when $\bQ$ is available it is expected to provide less efficient estimates comparing with $L^\star_Q$, as the latter exploits data available for cases with missing subtypes. 
\section{Simulation Study}
\label{Sec:Sims}
In the simulation study, we considered two possible subtypes. The baseline hazard for each subtype was of a Weibull distribution, with parameter values  chosen such that the baseline hazard ratio equals to $\alpha_2(t;\eta_1,\eta_2)=\eta_1 t^{\eta_2}$ with $\eta_1\simeq0.037$ and $\eta_2=1$.  We took $X$ as a single binary covariate with a prevalence of 0.4 and let the corresponding hazard ratios to be  $1.25$ and $2$ in subtypes $1$ and $2$, respectively.  Censoring was generated using an exponential distribution with mean 50, with additional type I censoring at time 90. The above parameter values were chosen  to have a censoring rate of about 70\%.  We took $Q$, the auxiliary variable, to be a single binary variable that we generated according to $P(Q=1|Y=1)=0.25$  and $P(Q=1|Y=2)=0.5$. 

We first considered the \sMARq scenario with a logistic regression model  $\pi(q)=\expit(\gamma_0+\gamma_qq)$. We considered various values for $\gamma_0$ and $\gamma_q$ determined by our desired values for $P(O=1|Q=0)$ and  $P(O=1|Q=1)$. In this scenario, we compared between three alternatives. The first was a complete case analysis (CCA), where cases with unknown subtype were thrown away and a standard competing risks model was used. Alternatives were estimates obtained from maximizing $L^\star_{Q2}$ in \eqref{Eq:LstarDecomp} and the estimates obtained under NMAR from maximizing $L^\star_{Y}$ while assuming the logistic regression model $\pi(y)=\expit(\tilde{\gamma}_0+\tilde{\gamma}_yI\{y=2\})$. Both estimation methods are theoretically valid if the truth is  \MARq; see details in Web Appendix B.

Summary of 1000 simulation iterations is presented in Table \ref{Tab:SimsMARq} for sample sizes $n=500,$ $1000,10000$ with approximately 150, 300, 3000 numbers of cases, respectively. Confidence intervals when using our methods were calculated using the asymptotic normal distribution  where variance estimates taken from a robust sandwich estimator, which is given in Web Appendix A. Both our methods showed good finite sample properties, with minimal bias even for a relatively small sample size, and desirable coverage rates for the corresponding confidence intervals. Whenever $Q$ played a significant role in the missing pattern, i.e., when $P(O=1|Q=0)$ and  $P(O=1|Q=1)$ were well differentiated, standard deviations of our estimators were higher, where bias was left unchanged. There was a large bias in the estimation by CCA, even when  $P(O=1|Q=1)=P(O=1|Q=0)=0.2$. This is because while the last equality stands, missing completely at random does not, since $P(O=1|Y=1)\ne P(O=1|Y=2)$. An efficiency gain is observed for $L^\star_Q$ comparing to $L^\star_Y$. This can be explained by the fact $L^\star_Q$ using more data than $L^\star_Y$, and specifically the data on $\bQ$ for observations with missing subtype.

We then turned to a more complex missingness model, which involves effects of  $X$, $T$, $Q$ and $Y$. First, we simulated the data under \sMARtxq  with the model $\pi(t, x, q)=\expit(\gamma_qq +0.5x - 0.01I\{t>50\})$, with values $\log(\gamma_q)\in  \{1,1.25,1.75,2.5,5\}$. Larger $\gamma_q$ value implies less cases with missing subtypes, and a more substantial role of $Q$ in determining $O$. Other than the missing model, we used the same values as in the previous scenario.  In addition to CCA and our methods, we also considered the estimating equations of \cite{goetghebeur1995analysis} (GR), which is valid under \MARtx. The $L^\star_Y$ method now estimates the parameters under the  model $\pi(t,x,y)=P(O=1|\tilde{T}=t,X=x,Y=y)=\expit(\gamma^*_yI\{y=2\}+\gamma^*_tI\{t>50\}+\gamma^*_xx)$. Note that, since $\pi(t, x, q, y)$, defined in \eqref{Eq:piDef}, can be written under \sMARtxq as 
$
\pi(t, x, q, y)= \pi(t, x, q=1)P(Q=1|\tilde{T}=t, X=x, Y = y) +  \pi(t, x, q=0)P(Q=0|\tilde{T}=t, X=x, Y = y),
$ the given model under \sMARtxq  with the logit function for $\pi(t, x, q)$  typically does not imply the logistic regression model for $P(O=1|\tilde{T}=t, X=x, Y=y)$. That is, this $L^\star_Y$ analysis used a misspecified missing data model for this simulation setup.

As a final scenario, we repeated the aforementioned simulations, but now taking the missing model assumed by $L^\star_Y$, $\expit(\gamma_yI\{y=2\}-0.01I\{t>50\}+0.5x)$, to be the correct model. That is, now $\gamma_q=0$. We  considered $\log(\gamma_y)\in  \{1, 1.25, 1.75, 2.5, 5\}$.  

Table \ref{Tab:SimsComplex} presents the results for both the \sMARtxq  and NMAR scenarios for $n=10,000$. While larger $\gamma_q$ or $\gamma_y$ values imply less missing subtypes, larger bias (in absolute value) was observed for estimates obtained by the GR method. For CCA, the substantial bias in estimation of both parameters was mildly reduced as $\gamma_q$ or $\gamma_y$ grew. For \MARtxq, performance of the $L^\star_{Q2}$ approach was not affected by the value of $\gamma_q$. While in this case $L^\star_{Y}$ used  a misspecified logistic model for the missing, it provides less biased estimates comparing to CCA and GR for most parameter values.  Looking at the results for NMAR,  $L^\star_{Y}$ had minimal bias while $L^\star_{Q}$ had small bias for $\gamma_y<2.5$, and generally lower bias than the bias observed for CCA and GR. Simulation results for lower sample size $n=1,000$, presented in Web Appendix C,  has shown larger, though moderate, bias for $L^\star_{Q}$ and $L^\star_{Y}$, but these methods remain superior to GR and CCA. 

The likelihood suggested by \cite{dewanji1992note} had been criticized for having a non mean-zero score function for $\bbeta$ if $\alpha_k$ is misspecified. To investigate this potential limitation, we considered the $L^\star_Q$ estimator when taking a piecewise constant function for the baseline hazard ratio $\alpha_k$ and the data was generated by the same model outlined above. The results, presented in Web Appendix C, did not reveal further bias when using piecewise constant function for the baseline hazard ratio $\alpha_k$.   

\section{Colorectal cancer analysis}
\label{Sec:Data}
To illustrate the use of our method, we present in this section an analysis of CRC data. CRC is regularly classified according to results of molecular studies \citep{ogino2011molecular,ogino2010lifestyle,campbell2010case,Kuipers2015}. One commonly-used classification is microsatellite instability (MSI) versus microsatellite stability (MSS). For example, high  body mass index (BMI) has been associated with MSS subtype of CRC tumors but not MSI subtype  \citep{hughes2012body}. 

We used the Nurses' Health Study (NHS) cohort, in which 121,701 female nurses enrolled in 1976 and since then answered  biennial  questionnaire regarding lifestyle and other risk factors. Whenever a new CRC cancer was diagnosed, the date was recorded as well as basic tumor characteristics,  available for all cases we considered. The tumor location is classified to be proximal colon, distal colon or rectum. Our analysis included 1,844 CRC cases that were observed, of which 121 MSI cases, 477 MSS cases, and the remaining cases (66\%) had missing MSI status. In total, our data include 114,073 participants, with 2,786,825 person-years of follow-up. Data on some risk factors were not measured from the beginning of the study, hence not all of the 121,701 participants were included in the data.

As a preliminary analysis,  we considered a logistic regression model for $O$, the indicator that MSI status was known, and we examined various covariates. Details are given in Web Appendix D. As presented in   Web Appendix D, missing MSI status  depends on the tumor location, even when including other covariates $(X)$ and the time of the diagnosis $(T)$. It was previously established that in CRC, MSI status and proximal location are associated \citep{colussi2013molecular}.  Thus, the traditional \sMARtx probably does not hold here.   We take here $Q$ to be the tumor location, equals to one for proximal tumors and zero otherwise. Based on subject-matter considerations, we assume that conditional on MSI status,  CRC tumor location is independent of the time of diagnosis and the measured risk factors. That is, we let  $P(Q=1|Y=k)=\expit(\psi_k), k=1,2$.  We consider the following risk factors: family history of CRC (binary), BMI (continuous),  regular aspirin  use (binary) and cumulative pack-year of smoking (continuous). To account for changes in the baseline hazard along the (calendar) years, we used a stratified version for the baseline hazard ratio $\alpha_2$. The stratified version, presented in detail in Web Appendix E, is a natural generalization of $L^\star_Q$ given by \eqref{Eq:LstarQ}.

Table \ref{Tab:Data} compares the results between using a  CCA with a stratified Cox model for the subtype-specific hazards, using the 598 cases with known MSI status and all CRC-free participants, and implementation of our method based on $L^\star_Q$ that uses data on all 1844 cases and all CRC-free participants. Differences are shown between the estimates in both methods. For example, for CRC MSI subtype the log-hazard ratio of aspirin is about 75\% larger (in absolute value) when using CCA comparing to using $L^\star_Q$.  The estimated standard deviations of the $L^\star_Q$-estimates, calculated using the robust sandwich estimator, are dramatically lower than those obtained by CCA. Furthermore,  the effect of aspirin on the incidence of MSS CRC  that was not significant  at $5\%$ level when using CCA is significant when using $L^\star_Q$. 

\section{Discussion}
\label{Sec:Discuss}
Existing methods for subtype analysis often rely on missing-at-random assumptions that may fail to hold in practice. Analysis under these assumptions potentially results in biased effect estimates. In this paper, we presented a weaker missing-at-random assumption which is reasonable when additionally  auxiliary covariates are measured for all cases. Under this assumption, we have developed an informative partial likelihood approach which results in consistent and normally distributed estimators. We also presented a method to conduct analysis when there are no auxiliary case covariates available and the  subtypes are NMAR, which is even a weaker assumption.


Comparing the two methods we have proposed, using $L^\star_Q$ alleviates the concerns of fitting the missing model $\pi$, but those are replaced with  questions regarding the model for $\bQ$.  Valid use of $L^\star_Y$ relies on correct specification of the missingness model but does not concern the distribution of $\bQ$. Furthermore, while  $L^\star_Q$ cannot be used when $\bQ$ is unavailable, $L^\star_Y$ can be used. Our simulation results suggested that estimates obtained by $L^\star_Q$ are more efficient than those obtained by $L^\star_Y$  in situations where both methods provide consistent estimates. Finally, both methods were shown to reduce bias of the estimators of interest comparing to the  traditionally used estimation equations of \cite{goetghebeur1995analysis}. 

Analysis by  $L^\star_Q$ offers additional two advantages. The first is that when the missing probability is a function of $\bX$, $T$ and $Y$, and the \sMARtxq does not hold, estimates using $L^\star_Q$ are expected to be less biased than estimates obtained under traditionally assumed \MARtx.  Second is that when \sMARtx holds, no damage is done by using $L^\star_Q$, which is still valid. Furthermore, one may expect an efficiency gain if $\bQ$ and $Y$ are associated, even if neither $\bQ$ or $Y$ are  associated with $O$ (conditionally on $T$ and $\bX$). Our simulation results in Section \ref{Sec:Sims} under NMAR and \sMARtx confirm these claims.

When $\bQ$ is high-dimensional, modeling the distribution of $\bQ$ may increase the total number of model parameters. It worth mentioning that from our point of view, $\bpsi$, the parameters characterizing the distribution of $\bQ$   are nuisance parameters, and exact correct specification of the joint distribution of $\bQ$ is not our goal. Thus, in order to get a low-dimensional vector or even a scalar $Q$ to be used in the main analysis,  a preliminary analysis of $\bQ$ yielding a working model to summarize $\bQ$ into low-dimensional vector (or scalar) $\bh(\bQ)$ can be used.  For example, one may apply a principal component analysis or a factor analysis.

When the number of subtypes of interest is relatively small, which is generally the case, using a piecewise constant function for the baseline hazard ratio $\alpha_k(t;\bbeta)$  can be a good alternative to a more restrictive parametric model. If the number of investigated subtypes is very large, our approach may result in a computational burden when estimating the large number of parameters.  Under the more restrictive \MARtx, the method suggested by \cite{chatterjee2010analysis} may serve as a potential good alternative. In their suggested method, the auxiliary covariates can be used as an additional biomarker, or ``disease trait'' in their terminology, that its crossproduct with other disease traits form the subtypes. From a clinical point of view, however, biomarkers (such as MSI) are usually used for subtype definition, while the other characteristics (such as tumor location) are treated as specific clinical properties of the disease. 


Analysis with our approach  will result in two major advantages for practitioners. First, potential bias will be eliminated. Second, data that are not used today can be utilized to get more efficient estimates, with better power for hypothesis testing and  narrower confidence intervals as a result of inclusion of cases that had been left aside so far. Hence our methods can be very useful in the era of MPE and precision medicine.

\section{Software}
\label{sec5}

R code implementing the methods presented in the paper is available on
request from the corresponding author (danielnevo@gmail.com).

\section{Supplementary Material}
\label{sec6}

Supplementary material is available online.
Web appendices A--E give technical details and proofs, present additional simulation results and contain more information on the CRC data example.

\section*{Funding}

This work was supported by National Institutes of Health grant numbers R35 CA197735, R01 CA151993, K07 CA190673, P50 CA127003, UM1 CA186107 and  P01 CA87969.


\bibliographystyle{plainnat}
\bibliography{missingsubtype}

\newgeometry{top=1in,bottom=1in,right=1in,left=1in}
\begin{table}[!p]
	\caption{Simulation results for complete case analysis (CCA) and our methods: $L^\star_Q$ under \sMARq  and $L^\star_Y$ under NMAR. True values for the parameters were $\beta_1=0.223=\log(1.25)$ and $\beta_2=0.916=\log(2.5)$.  Relative bias (\%) and standard deviation (SD) presented for both methods as well as confidence interval coverage rates  (CI-R) for our methods.
		\label{Tab:SimsMARq}}
	\small
	\centering
	{\footnotesize \tabcolsep=4.25pt
		\begin{tabular}{p{0.001cm}ccccccccccccc}
			\hline
			& 	\multicolumn{2}{c}{$P(O=1|Q)$} & \%Missing  & & \multicolumn{3}{c}{$n=500$} & \multicolumn{3}{c}{$n=1,000$} & \multicolumn{3}{c}{$n=10,000$}\\
				&	$Q=0$ &$ Q=1$	& Subtype& & CCA & $L^\star_Q$ & $L^\star_Y$ & CCA & $L^\star_Q$ & $L^\star_Y$ & CCA & $L^\star_Q$ & $L^\star_Y$ \\ \hline
			$\beta_1$ & 0.20 & 0.80 & 55\% & Bias(\%) & 46.65 & -7.20 & -8.99 & 58.36 & -3.48 & -3.90 & 57.86 & -0.58 & -0.54 \\ 
			&&& & SD  & 0.51 & 0.39 & 0.44 & 0.35 & 0.26 & 0.29 & 0.11 & 0.08 & 0.09 \\ 
			&&&	& CI-R  &  & 0.96 & 0.94 &  & 0.96 & 0.95 &  & 0.95 & 0.94 \\ 
			& 0.40 & 0.80 & 44\% & Bias(\%) & 32.60 & -7.46 & -10.98 & 40.43 & -5.65 & -4.66 & 43.23 & -0.44 & -0.73 \\ 
			&&& & SD  & 0.42 & 0.35 & 0.38 & 0.28 & 0.23 & 0.25 & 0.09 & 0.08 & 0.08 \\ 
			&&&	& CI-R  &  & 0.96 & 0.94 &  & 0.96 & 0.95 &  & 0.96 & 0.95 \\ 
			&0.80 & 0.80 & 20\% & Bias(\%) & 13.78 & -9.25 & -8.86 & 19.03 & -4.90 & -3.66 & 18.29 & -3.85 & -3.53 \\ 
			&&& & SD  & 0.32 & 0.31 & 0.32 & 0.22 & 0.21 & 0.21 & 0.07 & 0.07 & 0.07 \\ 
			&&&	& CI-R  &  & 0.95 & 0.95 &  & 0.95 & 0.95 &  & 0.94 & 0.94 \\ 
			&0.80 & 0.40 & 36\% & Bias(\%) & 39.53 & -4.28 & -2.99 & 43.03 & -0.88 & 0.51 & 43.05 & -0.59 & -0.08 \\ 
			&&& & SD  & 0.33 & 0.32 & 0.32 & 0.23 & 0.22 & 0.22 & 0.07 & 0.07 & 0.07 \\ 
			&&&	& CI-R  &  & 0.96 & 0.96 &  & 0.96 & 0.95 &  & 0.96 & 0.96 \\ 
			&0.80 & 0.20 & 45\% & Bias(\%) & 50.00 & -6.13 & -2.69 & 55.34 & -2.57 & -2.37 & 53.31 & -2.58 & -2.35 \\ 
			&&& & SD  & 0.35 & 0.33 & 0.34 & 0.25 & 0.23 & 0.24 & 0.08 & 0.07 & 0.08 \\ 
			&&&	& CI-R &  & 0.96 & 0.96 &  & 0.96 & 0.96 &  & 0.94 & 0.93 \\ 
			\hline
			$\beta_2$ & 0.20 & 0.80 & 55\% & Bias(\%) & 16.54 & 0.50 & 0.26 & 15.37 & -0.57 & -0.06 & 14.93 & -0.77 & -0.80 \\ 
			&&& & SD  & 0.31 & 0.25 & 0.27 & 0.22 & 0.18 & 0.19 & 0.07 & 0.06 & 0.06 \\ 
			&&&	& CI-R  &  & 0.96 & 0.95 &  & 0.95 & 0.95 &  & 0.94 & 0.94 \\ 
			& 0.40 & 0.80 & 44\% & Bias(\%) &13.61 & 1.06 & 0.94 & 12.87 & 0.17 & 0.35 & 11.26 & -0.85 & -0.86 \\ 
			&&& & SD  & 0.28 & 0.23 & 0.25 & 0.19 & 0.16 & 0.17 & 0.06 & 0.05 & 0.05 \\ 
			&&&	& CI-R  &  & 0.96 & 0.96 &  & 0.96 & 0.96 &  & 0.96 & 0.95 \\ 
			& 0.80 & 0.80 & 20\% & Bias(\%) & 7.55 & 1.20 & 1.61 & 5.71 & -0.62 & -0.58 & 5.28 & -0.66 & -0.65 \\ 
			&&& & SD  & 0.24 & 0.22 & 0.23 & 0.17 & 0.15 & 0.16 & 0.05 & 0.05 & 0.05 \\ 
			&&&	& CI-R  &  & 0.95 & 0.96 &  & 0.96 & 0.96 &  & 0.96 & 0.95 \\ 
			& 0.80 & 0.40 & 36\% & Bias(\%) & 12.01 & -0.06 & 0.86 & 10.85 & -1.12 & -1.15 & 11.24 & -0.53 & -0.50 \\ 
			&&& & SD  & 0.27 & 0.23 & 0.24 & 0.19 & 0.15 & 0.16 & 0.06 & 0.05 & 0.05 \\ 
			&&&	& CI-R  &  & 0.96 & 0.95 &  & 0.96 & 0.96 &  & 0.95 & 0.95 \\ 
			& 0.80 & 0.20 & 45\% & Bias(\%) & 16.63 & 0.49 & 1.57 & 14.85 & -0.74 & -0.30 & 14.55 & -0.54 & -0.51 \\ 
			&&& & SD  & 0.32 & 0.24 & 0.27 & 0.21 & 0.17 & 0.18 & 0.07 & 0.05 & 0.06 \\ 
			&&&	& CI-R  &  & 0.94 & 0.94 &  & 0.95 & 0.94 &  & 0.94 & 0.94 \\ 
			\hline
		\end{tabular}}
	\end{table}
	
	\begin{table}[!p]
					\centering
		\caption{Simulation results for complete case analysis (CCA), estimating equation approach (GR) and our methods: $L^\star_Q$  and $L^\star_Y$ when truth was NMAR. True values for the parameters were $\beta_1=0.223=\log(1.25)$ and $\beta_2=0.916=\log(2.5)$. Sample size for each simulation iteration was 10,000.}
			\label{Tab:SimsComplex}
	
		{	\footnotesize \tabcolsep=4.25pt
			\begin{tabular}{cccccccccccc}	
				\hline
						&		{$(e^{\gamma_q},e^{\gamma_y})$}	& \%Missing & &	\multicolumn{4}{c}{$\beta_1$ (true value $0.223$)} & 	\multicolumn{4}{c}{$\beta_2$ (true value $0.916$)}\\
					&		& Subtype	& & CCA & $L^\star_Q$ & $L^\star_Y$ &  GR & CCA & $L^\star_Q$ & $L^\star_Y$ & GR \\ \hline
				\multirow{10}{*}{\MARtxq} & (1,0) & 49.5\% & Bias(\%) &  122.65 & -2.02 & -1.37 & -1.27 & 33.06 & -0.69 & -0.73 & -0.63 \\ 
				&		&& SD & 0.09 & 0.08 & 0.08 & 0.08 & 0.07 & 0.05 & 0.05 & 0.06 \\ 
				&		(1.25,0) & 47.2\% & Bias(\%) & 115.70 & -2.28 & -0.95 & -5.57 & 30.71 & -0.65 & 0.39 & -2.27 \\ 
				&		&& SD 	& 0.08 & 0.07 & 0.07 & 0.08 & 0.07 & 0.05 & 0.05 & 0.07 \\ 
				&		(1.75,0) &  44.1\% & Bias(\%) & 108.93 & -0.85 & 2.13 & -11.85 & 27.12 & -0.92 & 1.67 & -7.06 \\ 
				&		&& SD 	& 0.08 & 0.07 & 0.07 & 0.08 & 0.06 & 0.05 & 0.06 & 0.07 \\ 
				&		(2.5,0) & 41.0\% & Bias(\%) & 98.95 & -1.30 & 4.93 & -15.61 & 23.58 & -1.10 & 2.79 & -7.64 \\ 
				&		&& SD 	& 0.08 & 0.07 & 0.07 & 0.08 & 0.06 & 0.05 & 0.06 & 0.06 \\ 
				&		(5,0) & 36.1\% & Bias(\%) & 88.04 & -0.93 & 12.20 & -21.18 & 19.02 & -0.82 & 4.68 & -9.45 \\ 
				&		&& SD 	& 0.08 & 0.07 & 0.07 & 0.08 & 0.06 & 0.05 & 0.06 & 0.06 \\ 
				\multirow{10}{*}{NMAR} &	(0,1) & 49.5\% & Bias(\%) & 122.41 & -1.63 & -1.16 & -1.04 & 32.88 & -0.79 & -0.84 & -0.72 \\  
				&	&& SD 			& 0.09 & 0.07 & 0.07 & 0.08 & 0.07 & 0.05 & 0.05 & 0.07 \\ 
				&	(0,1.25) & 46.1\% & Bias(\%) & 117.19 & -2.50 & -0.13 & -6.34 & 28.07 & -3.16 & -1.17 & -5.40 \\ 
				&	&& SD & 0.09 & 0.08 & 0.08 & 0.08 & 0.06 & 0.05 & 0.06 & 0.07 \\ 
				&	(0,1.75) &  41.0\% & Bias(\%) & 108.24 & -3.99 & -1.56 & -15.34 & 22.93 & -5.21 & -0.47 & -9.97 \\ 
				&	&& SD 	& 0.09 & 0.08 & 0.07 & 0.08 & 0.06 & 0.05 & 0.05 & 0.06 \\ 
				&	(0,2.5) & 36.2\%  & Bias(\%) & 101.96 & -1.12 & -0.33 & -20.44 & 17.35 & -7.49 & -0.48 & -14.45 \\ 
				&	&& SD  & 0.08 & 0.08 & 0.07 & 0.08 & 0.06 & 0.05 & 0.05 & 0.06 \\ 
				&	(0,5) & 28.7\% & Bias(\%) & 93.74 & 2.59 & -1.28 & -31.59 & 9.44 & -10.44 & -0.62 & -22.55 \\ 
				&	&& SD 			& 0.09 & 0.08 & 0.07 & 0.08 & 0.05 & 0.04 & 0.05 & 0.06 \\ 
				\hline			
			\end{tabular}}
		\end{table}
		
		\begin{table}[!p]
			
			Risk factor analysis of CRC according to MSI status using a complete case analysis (CCA) and our proposed method $L^\star_Q$. The examined risk factors are 
				family history of CRC (Family), BMI,  regular aspirin  use (Aspirin) and cumulative pack-year of smoking (SmokePkyr).
				\label{Tab:Data}
				\centering
			{\tabcolsep=4.25pt
				\begin{tabular}{ccccc|ccc}	
					\hline
						& &	\multicolumn{3}{c}{CCA} &	\multicolumn{3}{c}{$L^\star$}   \\
						& &	$\hat{\beta}(e^{\hat{\beta}})$ & $\hat{SD}(\hat{\beta})$ &  $p$ & $\hat{\beta}(e^{\hat{\beta}})$ & $\hat{SD}(\hat{\beta})$  & $p$ \\ \hline
					\multirow{5}{*}{MSI} & BMI 	     & 0.013(1.013)  & 0.017 & 0.47 	& 0.012(1.012)  & 0.015 &   0.431\\
					& Aspirin   & -0.279(0.757) & 0.196 & 0.15     & -0.168(0.845) & 0.149 &   0.26\\
					& SmokePkyr & 0.013(1.013)  & 0.003 & $<0.001$ & 0.014(1.014)  & 0.002 &  $<0.001$\\
					& Family    & 0.738(2.091)  & 0.151 & $<0.001$ & 0.605(1.832)  & 0.137 &  $<0.001$\\
					\multirow{5}{*}{MSS} & BMI 	     & 0.026(1.026)  & 0.008 & 0.002 	& 0.018(1.018)  & 0.006 &  0.001 \\
					& Aspirin   & -0.186(0.831) & 0.100 & 0.063    & -0.251(0.778) & 0.063 &  $<0.001$ \\
					& SmokePkyr & 0.006(1.006)  & 0.002 & $0.002$  & 0.007(1.007)  & 0.001 &  $<0.001$ \\
					& Family    & 0.274(1.316)  & 0.099 & $0.006$  & 0.200(1.221)  & 0.072 & 0.005 \\
					\hline
				\end{tabular}}
			\end{table}

		\end{document}